\documentclass[aps,prl,preprint,nopacs,superscriptaddress]{revtex4}

\usepackage{graphicx}
\usepackage{verbatim}
\usepackage{mathrsfs}
\usepackage{gensymb}
\usepackage{ulem}

\usepackage{amsmath,amsfonts,amssymb}
\usepackage{graphicx}
\def\3{2.8in}    
\def\2{2.5in}
\def\4{3.0in}\def \beq {\begin{equation}}
\def \eeq {\end{equation}}
\begin{document}

\title{Engineering electronic structure of a 2D topological insulator Bi(111) bilayer on Sb nanofilms by quantum confinement effect}

\author{Guang~Bian\footnote{These authors contributed equally to this work.}}
\affiliation {Joseph Henry Laboratory, Department of Physics, Princeton University, Princeton, New Jersey 08544, USA}

\author{Z. F. Wang$^*$}
\affiliation {Department of Materials Science and Engineering, University of Utah, Salt Lake City, Utah 84122, USA}

\author{Xiaoxiong Wang}
\affiliation {College of Science, Nanjing University of Science and Technology, Nanjing 210094, China}

\author{Caizhi Xu}
\affiliation {Department of Physics, University of Illinois at Urbana-Champaign, 1110 West Green Street, Urbana, Illinois 61801-3080, USA}
\affiliation {Frederick Seitz Materials Research Laboratory, University of Illinois at Urbana-Champaign, 104 South Goodwin Avenue, Urbana, Illinois 61801-2902, USA}

\author{Su-Yang~Xu}
\affiliation {Joseph Henry Laboratory, Department of Physics, Princeton University, Princeton, New Jersey 08544, USA}

\author{T. Miller}
\affiliation {Department of Physics, University of Illinois at Urbana-Champaign, 1110 West Green Street, Urbana, Illinois 61801-3080, USA}
\affiliation {Frederick Seitz Materials Research Laboratory, University of Illinois at Urbana-Champaign, 104 South Goodwin Avenue, Urbana, Illinois 61801-2902, USA}

\author{M. Zahid Hasan}
\affiliation {Joseph Henry Laboratory, Department of Physics, Princeton University, Princeton, New Jersey 08544, USA}
\affiliation {Princeton Center for Complex Materials, Princeton University, Princeton, New Jersey 08544, USA}

\author{Feng Liu}
\affiliation {Department of Materials Science and Engineering, University of Utah, Salt Lake City, Utah 84122, USA}
\affiliation {6 Collaborative Innovation Center of Quantum Matter, Beijing 100084, China}

\author{T.-C. Chiang}
\affiliation {Department of Physics, University of Illinois at Urbana-Champaign, 1110 West Green Street, Urbana, Illinois 61801-3080, USA}
\affiliation {Frederick Seitz Materials Research Laboratory, University of Illinois at Urbana-Champaign, 104 South Goodwin Avenue, Urbana, Illinois 61801-2902, USA}

\pacs{}

\date{\today}

\begin{abstract}
We report on fabrication of a two-dimensional topological insulator-Bi(111) bilayer on Sb nanofilms via a sequential molecular beam epitaxy (MBE) growth technique. Our angle-resolved photoemission measurements demonstrate the evolution of the electronic band structure of the heterostructure as a function of the film thickness and reveal the existence of a two-dimensional spinful massless electron gas within the top Bi bilayer. Interestingly, Our first-principles calculation extrapolating the observed band structure shows that, by tuning down the thickness of the supporting Sb films into the quantum dimension regime, a pair of isolated topological edge states emerges in a partial energy gap at 0.32 eV above the Fermi level as a consequence of quantum confinement effect. Our results and methodology of fabricating nanoscale heterostructures establish the Bi bilayer/Sb heterostructure as a platform of great potential for both ultralow-energy-cost electronics and surface-based spintronics. 
\end{abstract}

\maketitle

The last decade has witnessed an enormous research interest in exploration of  various topological electronic phases of condensed matters \cite{RMP, Zhang_RMP}. Many distinct topological phases protected by either the time reversal symmetry or crystalline symmetries have been experimentally realized. Among these emergent phases, the two dimensional (2D) topologial (quantum spin Hall) insulator (TI), the first-ever proposed prototype of topological insulators (TI), stands uniquely important for nanoelectronics applications. This system hosts one dimensional topological edge modes in the energy gap with spin-up and spin-down carriers propagating in opposite directions \cite{Kane_PRL,Zhang_PRL,Zhang_Science, Molenkamp_Science,RRDu_1,RRDu_2, Moler}. The spin-momentum-locked character, arising as a consequence of the time-reversal symmetry, suppresses backscatterings in the edge conduction channels, which is a key feature for the realization of dissipationless electronic devices. One of prototypical 2D TIs is Bi(111) bilayer and there have been significant experimental progresses to prepare this artificial material in laboratory \cite{Murakami_PRL,Murakami_Science,Wada_PRB,Bansil_APL,FLiu_NComm,Hirahara,Dong,Yeom_PRB,Yazdani,Eich,FLiu_PRB,JiFeng_PRB}. The topological edge modes have been observed on a Bi(111) bilayer prepared on Bi$_2$Te$_3$ and bulk Bi. The edge modes are exposed on the surface, which permits direct spectroscopic characterizations and surface manipulation at the atomic level. However, in these experiments, the substrates supporting Bi(111) bilayer are metallic and, thus, the edge modes in these systems are degenerate with the bulk modes. As a result, their contribution to the system conductivity is overwhelmed by the bulk conductivity, making them nonideal for $\textit{isolated}$ conduction channels needed for 2D TI applications.

One way to make progress toward useful isolated 2D TI is to employ ultrathin films as substrates, instead of directly growing 2D TI overlayer on a semiconductor substrate as theoretically proposed \cite{Zhou_PNAS, Miao_PNAS}, but could be experimentally challenging.  The thinner the supporting substrate is, the fewer bulk carriers contribute to the overall conductance. More precisely, when the thickness of the substrate is reduced to a few atomic layers, the valence and conduction bands of the substate are discretized into quantum well (QW) subbands as a result of quantum confinement \cite{Tai}. The energy separation between adjacent QW subbands increases as the substrate film thickness is reduced. The resulting energy gaps can be made large enough to suppress or even eliminate the bulk conduction channels making topological edge modes in 2D TI more pronounced. This idea motivated us to fabricate by the method of molecular beam epitaxy (MBE) Bi(111) bilayer on ultrathin Sb films, another topological system with a similar crystal structure \cite{GB_Sb_1,GB_Sb_2}. In our experiment, the thickness of the supporting Sb films can be precisely controlled down to four atomic layers. We perform a systematic angle-resolved photoemission measurement (ARPES) on the MBE-grown heterostructure with various Sb film thicknesses, demonstrating the evolution of the electronic band structure of the heterostructure as a function of the film thickness and revealing the existence of a 2D spinful massless electron gas within the top Bi bilayer. All these experimental observations are reproduced by our first-principles calculation with a remarkable consistency. Interestingly, our calculation further show that there exist a pair of isolated topological edge states at the edge of Bi bilayer within a partial bulk band gap about 0.32 eV above the Fermi level. The emergence of the partial bulk band gap and the isolation of the topological edge states from the bulk states is a direct consequence of quantum confinement effect in the reduced dimension as discussed above.  Within this partial gap, even though there still exist bulk state residuals, the contribution from the topological edge modes can dominate in conduction, forming a spin channel with backscattering suppressed. Therefore, we show through a combined method of ARPES experiment and first-principles calculation that the adverse hybridization effect of the substrate on the functional overlayer can be systematically suppressed by reducing the thickness of the substrate, and thus, the desirable properties of overlayer can be enhanced by the quantum size effect (QSE) in the hybrid system.


The Bi/Sb samples were fabricated $\textit{in situ}$ by a sequential MBE deposition procedure as schematically illustrated in Figure 1. The details are elaborated in the Methods section. Most importantly, the structure of the ``add-on" film in each step is optimized by annealing to an appropriate temperature to ensure film smoothness. The structural quality was examined by RHEED and ARPES. Both Sb and Bi share a similar rhombohedral structure, which consists of slightly buckled honeycomb atomic layers stacked along the (111) direction. We follow the convention of calling the buckled layer a bilayer (BL) \cite{Liu_Allen,Koroteev,GB_Bi,GB_Sb_1,GB_Sb_2}.  On the (111) surface, the lattice constant of Sb is 5$\%$ smaller than that of Bi. The Bi(111) BL deposited on the Sb(111) surface adopts the Sb lattice constant  and is thus laterally compressed, as verified by the unchanged RHEED spot positions before and after the Bi BL growth. The sizable strain makes our Bi BL film quantitatively different from the BLs studied previously, but this does not alter the nontrivial topological order according to our calculation. The atomic structure of the resulting Bi/Sb(111) system  (Fig. 2a) is determined theoretically by optimizing the atomic coordinates. The vertical buckling within the Bi BL is found to be 1.739 and 1.742 $\textrm{\AA}$ for 4 BL and 20 BL Sb substrates, respectively, and the corresponding spacing between the Bi BL and the Sb surface layer is 2.621 and 2.600 $\textrm{\AA}$, respectively. By comparison, the intra- and inter-BL distances for bulk Sb(111) are 1.510 and 2.240 $\textrm{\AA}$, respectively. The primitive lattice vector of the surface unit cell is 4.307 $\textrm{\AA}$, and the SBZ is shown in Fig. 2(b).

 The electronic band structure of the Bi/Sb hetetrostucture during the process of atomic layer fabrication was systematically recorded by ARPES. Figs. 2(c) and 2(d) present the APRES spectra taken from 4- and 20 BL Sb(111) epitaxial films (before Bi growth). Right around the Fermi level are two Rashba topological surface bands that cross to form a Dirac cone at the SBZ center. The other bands at higher binding energies are QW subbands; these are well resolved and correspond to a definitive film thickness. This atomic-level film thickness uniformity is a prerequisite for the fabrication of precision quantum device structures. After Bi BL deposition and subsequent annealing, the spectra changed dramatically [Figs. 2(e) and 2(f)]. The topological electron-like surface states of Sb disappear while a pair of hole-like Rashba bands emerges in the Sb valence band region. These newly emerged bands are derived from the Bi states, which however hybridize with the Sb QW subbands as evidenced by kinks where they cross; an example of the kink is indicated by the arrow in each case. The hybridization makes the Bi-derived states surface resonance states. 
 
 To further explore the nature of the Bi-derived Rashba states, we mapped the electronic bands with different photon energies. As seen in the results [Figs. 3(a-c) and 3(g-i) for the 4- and 20 BL Sb substrates, respectively], the Sb QW subbands do not disperse as a function of photon energy but undergo strong intensity variation. The subbands near the valance band maximum are more intense at 22 eV, but this maximum intensity region shifts toward higher binding energies as the photon energy increases to 28 eV. This is typical for direct band transitions in the bulk \cite{Tai}. On the other hand, no prominent intensity variations are found for the Bi Rashba states as the photon energy varies; this behavior is consistent with the surface character of these states. 
 
 Our first-principles calculations of the band structure [Figs. 3(d) and 3(j)] are in good agreement with the main features of the ARPES data. The gray scale of the plotted bands indicates the projection of each state on the Bi 6$\textit{p}$ orbitals, which should simulate the ARPES intensity because of a short photoelectron mean free path. As expected, the Bi-derived Rashba states (labeled Bi1) are the most intense ones in the ARPES data. The calculation also reveals another set of Bi-derived states (labeled Bi2) at about 0.32 eV above the Fermi level near the zone center. States Bi1 and Bi2 would be degenerate if the Bi BL were freestanding  \cite{Dong,Yeom_PRB}, but become split in the present case because of interaction with the substrate. As shown in both ARPES spectra and theoretical band structure, the outer part (farther away from $\bar{\Gamma}$) of Bi1 band strongly hybridizes with the Sb bulk state, losing its surface character, while the inner part remains tightly localized within the Bi overlayer as evidenced by the Bi atomic orbital projection in Figs. 3(d) and 3(j). The band structure around the central crossing point of Bi1 band resembles a massless 2D electron gas (2DEG) as denoted by the arrow in Fig.~3(j). The calculated spin texture of the band structure [Figs. 3(f) and 3(k)] indicates a very high degree of spin polarization close to 90\% for states Bi1 because of a very strong spin-orbit coupling in Bi. This spinful 2DEG on the surface of this hybrid structure can be an ideal platform for future surface-based spintronics.

 A key feature of the calculated band structure for the case of the 4 BL Sb substrate is an energy gap of about 36 meV in the band structure along the $\bar{\Gamma}\--\bar{M}$ direction as highlighted in the zoom-in figure [Fig. 3(e)], where the gap is 0.32 eV above the Fermi level as marked by two horizontal dotted lines. This gap lies very close to the Bi-derived states Bi2 which also shows strong spin polarization, and is consistent with the TI band gap of topmost Bi bilayer as reported in \cite{Dong}. However, unlike all previous works, in our hybrid structure with 4BL Sb substate there is no Sb bulk state lying with this energy gap at least along $\bar{\Gamma}\--\bar{M}$, reflecting the suppression on bulk carrier by the quantum confinement effect. By contrast, the band structure for the 20 BL Sb substrate [Fig. 3(k)] shows no such gap because the Sb QW subbands are more densely populated. Thus, the system is metallic and cannot support isolated edge conduction channels, and the same is true for even thicker Sb substrates. So we see that decresing the Sb-substrate thickness can reduce the hybridization between the functional overlayer and the substrate and, potentially, achieve an approximation of the desired properties of the freestanding overlayer system which, otherwise, cannot exist alone in real materials. 
       
  The complete electronic band structure of the Bi BL/4-BL Sb(111) heterostructure is plotted in Fig. 4(a). A gap, seen along the $\bar{\rm{\Gamma}}-\bar{\textrm{M}}$  direction, arises from quantum confinement. We note that this is not an absolute gap, but a partial (directional) energy gap, as its energy range is traversed along the $\bar{\rm{\Gamma}}-\bar{\textrm{K}}$  direction by two bands which are the Rashba bands of the Sb(111) surface \cite{GB_Bi, GB_Sb_1, GB_Sb_2, Sasaki, Sasaki2, Ishida}. However, these Sb bands do not couple in $k$ space to Bi2 band along the $\bar{\rm{\Gamma}}-\bar{\textrm{M}}$ direction, which enable topological edge states to form within the gap. These edge states are further isolated from the other states by their spatial localization. The existence of the topological edge states in the partial energy gap opened by the quantum confinement effect is confirmed by direct calculation for a Bi BL nanoribbon made of 8 zigzag atomic chains on a 4 BL Sb substrate; an end cross-section view of the atomic structure is presented in Fig. 4(f). The overall surface-weighted band structure and spin texture [Figs. 4(b) and 4(c), respectively] reveals a partial energy gap as marked by the dashed lines. This gap, about 36 meV, is larger than the infinite-film case due to lateral quantization. The topological edge states are revealed by computing the edge-weighted band structure [Fig. 4(d)]. There is just one pair (an odd number) of edge states, as denoted by ``A", within the gap on each edge, in agreement with the two-dimensional topological order of the Bi(111) BL. (The tiny splitting for each edge band is an artifact arising from the finite width of the ribbon and the resulting coupling between the two edges.) The calculated spin texture [Figs. 4(e,f)] confirms that this pair of edge states has their spin direction locked to the momentum direction in a chiral spin configuration. The calculated charge density of a typical in-gap edge state A [indicated in Fig. 4(d)] reveals that this state is indeed localized at the edge of the Bi ribbon within a decay length of about 2 atomic chains [Fig. 4(g)]. We note that even thought there still exist bulk state residuals with this energy range due to the partial opening nature of the gap, the edge states are isolated in both real and momentum spaces as shown in Figs. 4(f) and 4(g), the scattering between the edge modes and the substrate bulk states is largely suppressed. 
 
 Even though the topological edge spin channel is at 0.32 eV above the Fermi level, one can make use of the protected edge conducting channel of the heterostructure by employing a gate bias in a device structure as depicted in Fig. 4(h). The proposed structure involves a top metal gate, which is separated from the heterostructure by a high-k dielectric layer (such as HfO$_2$). For the Bi BL/4-BL Sb(111) system, a shift of the Fermi level by 0.32 eV corresponds to a charge accumulation of 0.6 electrons per surface unit cell or 3.8 $\times 10^{14}$ electrons$/\rm{cm}^2$. The required gate voltage is about 3 V for a dielectric thickness of 10 $\rm{\AA}$, which is well within the usual device operating range. By tuning the gate voltage one can switch into and out of the energy range where the topological edge modes contribute importantly and, thus, achieve a topological-edge-mode-based transistor. By contrast, the same Fermi level tuning for the Bi BL/20-BL Sb(111) system would require a charge accumulation of about 4.5 electrons per surface unit cell. The required gate voltage would be ~30 V, and dielectric breakdown would become a concern. The use of nanofilms is an important enabling feature for this application. 

While 2D TIs have been realized in several cases in the past, none has been found to simultaneously possess a sizable gap (relative to the thermal energy at room temperature) and edge modes that are electrically isolated from parallel non-topological conduction channels. Our work shows a novel route to overcome the bottleneck thanks to the quantum confinement effect intrinsic to the quantum-well nanofilms. The system Bi/Sb is particularly promising because it can be made with atomic precision on standard Si substrates as demonstrated by our experiment, which enables tuning the Fermi level of the nanofilm readily and realizing novel devices like a topological-edge-mode-based transistor. As a possible further application of this 2D TI hybrid system, it would be interesting to synthesize it on top of a superconductor. Because the total thickness of the Bi BL plus 4 BL Sb, less than 2 nm, can be much less than the superconducting coherent length for a properly chosen superconducting substrate, the topological edge modes could form Cooper pairs as a consequence of proximity coupling. Thus, this system is a promising candidate for realizing one-dimensional topological superconductors \cite{Kane_Proximity,TSC_Xu,Zhang_TSC,Linder,Patrick_Lee}. We believe the methodology explored here offers a powerful approach to fabricate and control various nanoscale functional structures relevant to the development of next generation electronic devices.

\section{Methods}
The Bi BL/Sb(111) heterostructures were prepared through a sequential MBE deposition. An n-type Si(111) wafer (Sb-doped with a resistivity of ~0.01 $\Omega\cdot$cm) was cleaned by direct current heating to yield a (7$\times$7) reconstructed surface. It was cooled to 60 K, and about 6 $\textrm{\AA}$ of Bi was deposited on top. The sample was then annealed at 600 K for 10 minutes to yield a well ordered Bi-$\sqrt{3}\times\sqrt{3}$ surface reconstruction. This step was key to the growth of smooth Sb films. The deposition of Sb was performed with the substrate at 60 K with the  deposition rate monitored by a quartz thickness monitor. The resulting structure was annealed at 500 K to yield an atomically uniform Sb film. Finally, an appropriate amount of Bi was deposited on the surface and the system was annealed to 450 K for 5 minutes to yield the final sample. RHEED was employed to monitor the surface condition during the growth process. ARPES measurements were performed at the Synchrotron Radiation Center, University of Wisconsin-Madison. A Scienta analyzer equipped with a two-dimensional detector was employed for data collection. The overall energy resolution was 15 meV.

DFT calculations for Bi(111) BL on Sb(111) thin films are carried out in the framework of the Perdew-Burke-Ernzerhof-type generalized gradient approximation using the Vienna Ab initio simulation package. The lattice constant of Sb thin films is taken from its bulk value a=4.307 {\AA}, and Bi BL is strained to match the Sb lattice constant. All calculations are performed with a plane-wave cutoff of 400 eV. A 11$\times$11$\times$1 and 1$\times$11$\times$1 Monkhorst-Pack k-point mesh is used for bulk and ribbon calculations, respectively. The Sb thin films are modeled by a slab of 4 BL and 20 BL Sb, and the vacuum layers are over 15 {\AA} thick to ensure decoupling between neighboring slabs. For Bi BL on 4 BL Sb, all atoms are allowed to relax until the forces are smaller than 0.01 eV/{\AA}. For Bi BL on 20 BL Sb, the Bi BL and upper 6 BL Sb  are allowed to relax until the forces are smaller than 0.01 eV/{\AA}, while atoms in the lower 14 BL Sb  are fixed in their respective bulk positions.

\section{Acknowledgements}

This work was supported by the U.S. Department of Energy (DOE), Office of Science (OS), Office of Basic Energy Sciences, under Grant No. DE-FG02-07ER46383 (TCC), DE-FG02-04ER46148 (FL, ZFW),  NSF-MRSEC-Grant No. DMR-1121252 (ZFW), the National Science Foundation of China under Grant No. 11204133 (XW), the Jiangsu Province Natural Science Foundation of China under Grant No. BK2012393 (XW), and the Young Scholar Project of Nanjing University of Science and Technology (XW). Work at Princeton is funded in part by the Gordon and Betty Moore Foundation's EPiQS Initiative through Grant GBMF4547. The theoretical work is conducted at University of Utah using the CHPC and NERSC computing resources. We thank M. Bissen and M. Severson for assistance with beamline operation at the Synchrotron Radiation Center, which was supported by the University of Wisconsin-Madison. TM and the beamline operations were partially supported by NSF Grant No. DMR 13-05583.

\newpage

\begin{figure}
\centering
\includegraphics[width=11cm]{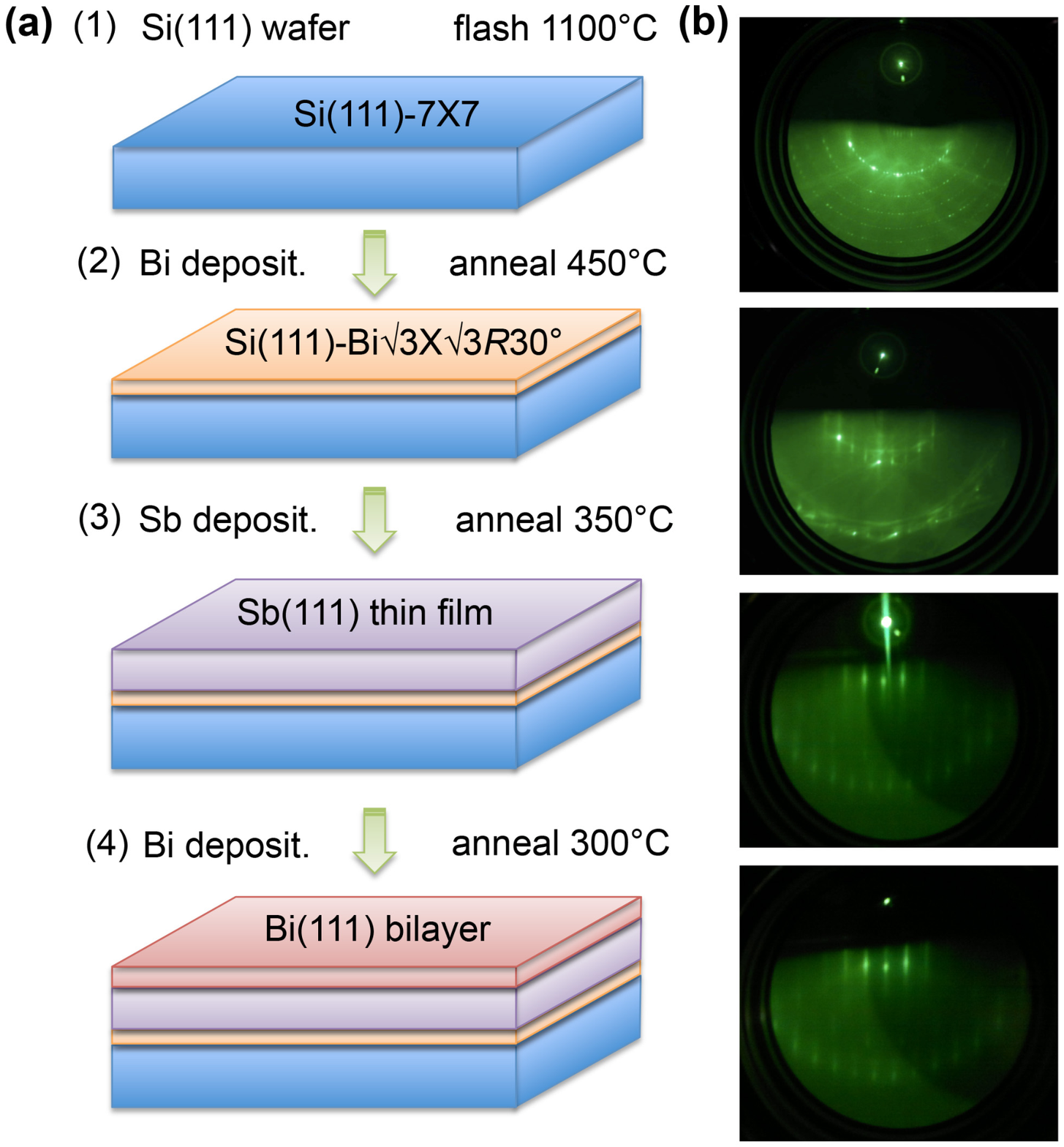}
\caption{(a) Sequential MBE fabrication of Bi BL/Sb(111) heterostructure. Schematic of the sample construction after each deposition step. (b) RHEED pattern taken after each deposition and annealing step.}
\end{figure}

\newpage

\begin{figure}
\centering
\includegraphics[width=14cm]{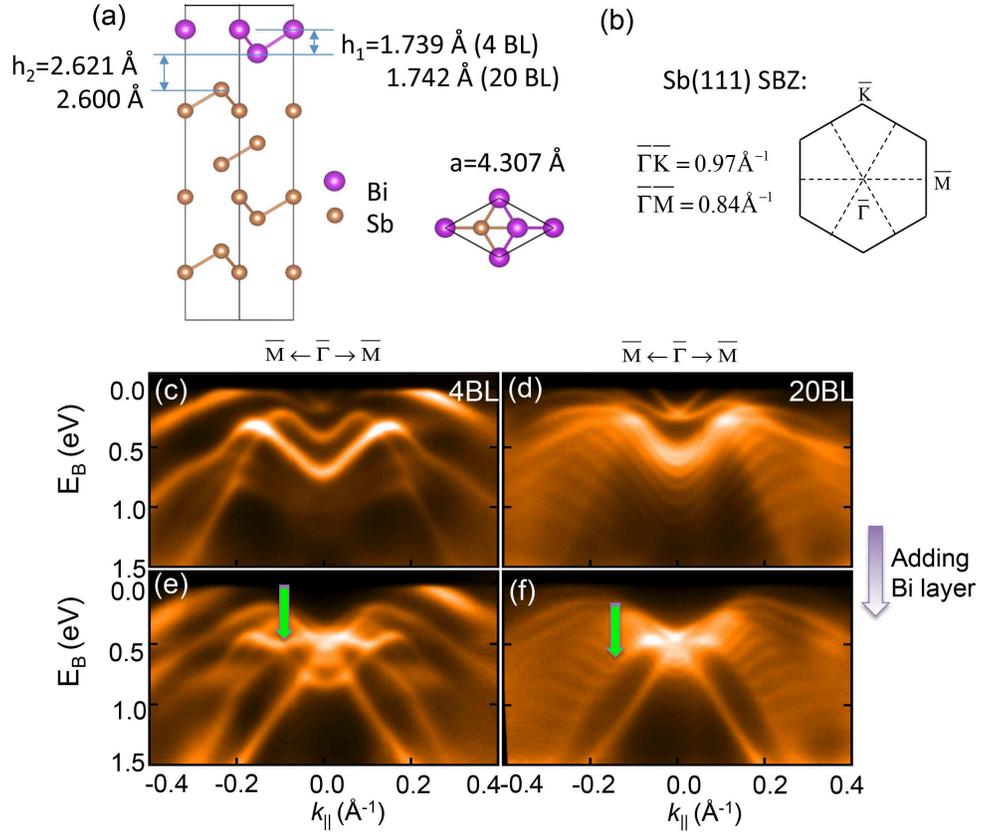}
\caption{(a) Side view and top view of the lattice structure of Bi BL/Sb(111) heterostructure. (b) The surface Brillouin zone.  ARPES data of (c) 4 BL Sb film, (d) 20 BL Sb film, (e) Bi BL on a 4 BL Sb film and (f) Bi BL on a 20 BL Sb film.} 
\end{figure}

\newpage

\begin{figure}
\centering
\includegraphics[width=16cm]{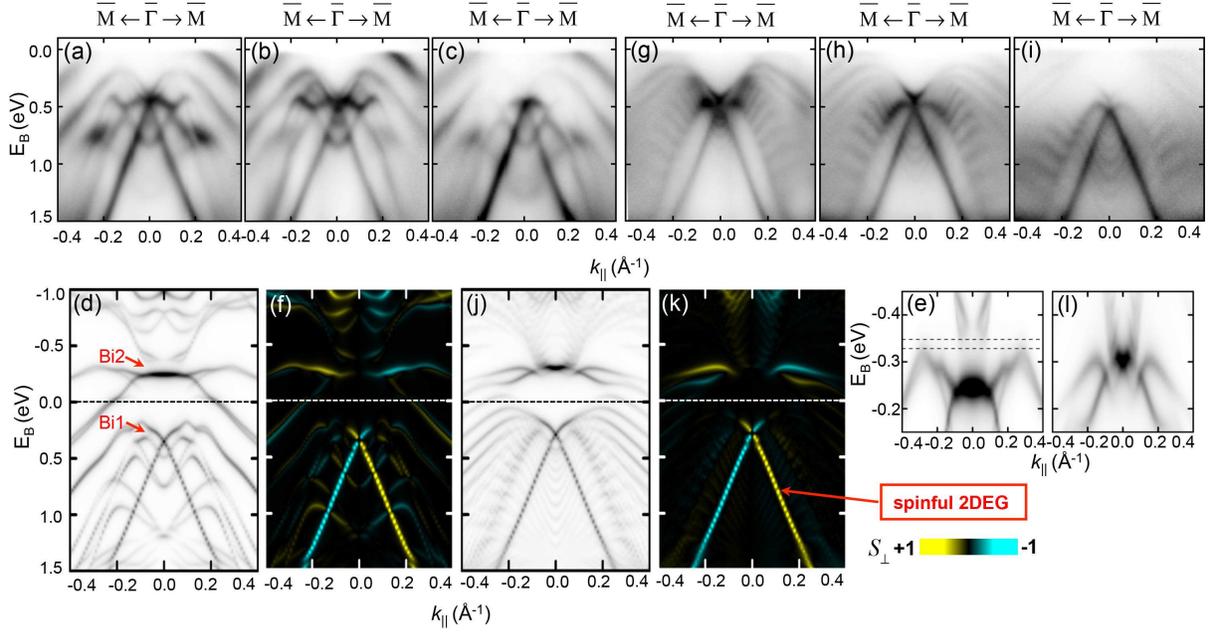}
\caption{(a-c) ARPES spectra taken from Bi BL/4BL Sb(111) with  22, 25, and 28 eV photons. (d) Surface projected DFT band structure of Bi BL/4BL Sb(111). The grey scale indicates the weight of the wavefunction within the surface layer. (e) Zoom-in view of DFT bands around 0.3 eV above the Fermi level. (f) Calculated spin polarization of states along the direction perpendicular to both the corresponding momentum and the surface normal. (g-l) The corresponding plots for Bi BL/20BL Sb(111).}
\end{figure}

\newpage

\begin{figure}
\centering
\includegraphics[width=16cm]{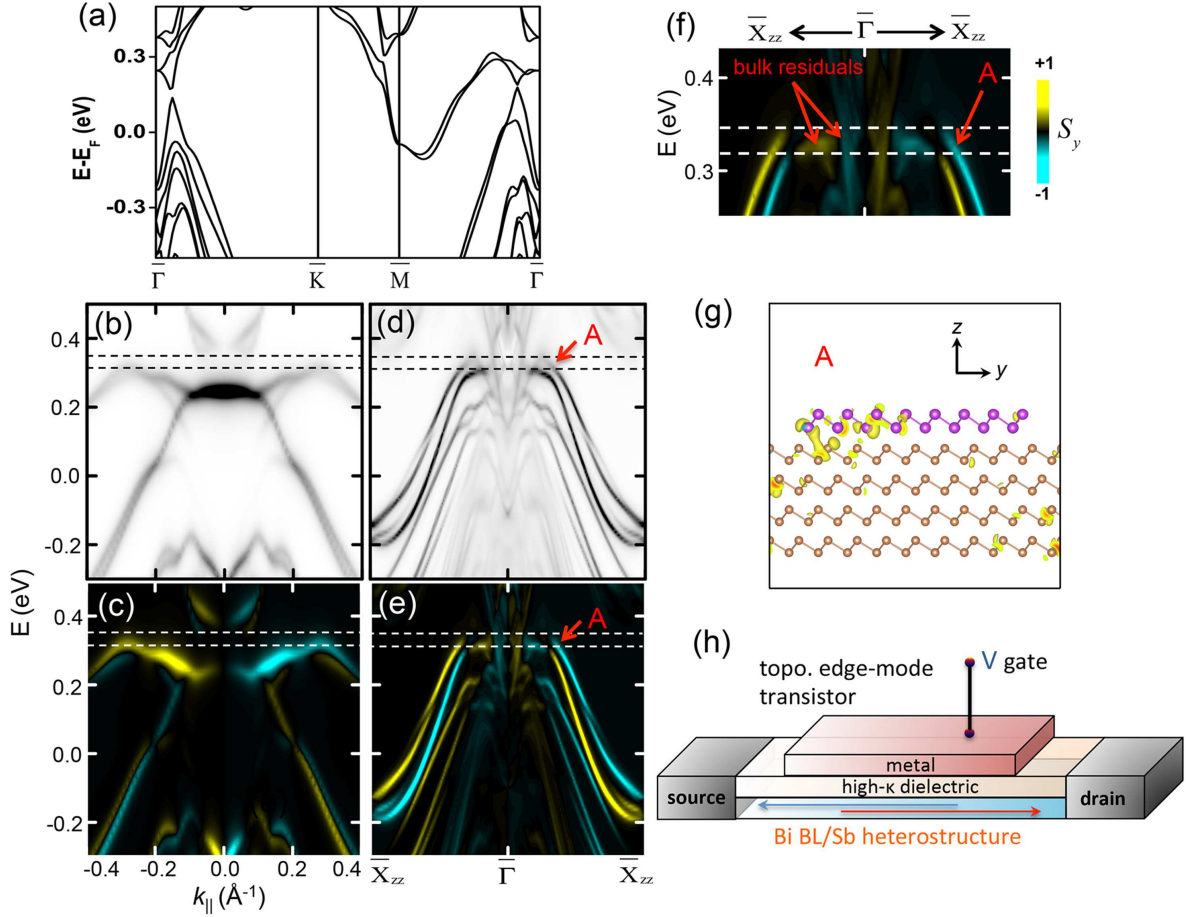}
\caption{(a) Bulk structure of Bi BL/4BL Sb(111). (b,c) Surface projected band structure and spin polarization of Bi BL/4BL Sb(111).  (d,e) Edge projected band structure and spin polarization of a Bi BL nanoribbon made of 8 atomic chains along the armchair direction on a 4 BL Sb substrate. The grey scale indicates the weight of the wavefunction on the edge atomic chain of Bi Bl. (f) Zoom-in spin map close to the partial energy gap (top). (g) The charge density distribution of edge state A indicated with the arrow (bottom). (h) Schematic of a gate controlled device based on Bi BL/Sb(111) heterostructure of a nanoribbon shape which can be switched into the topological edge-mode conducting regime.}
\end{figure}

\end{document}